\let\ifarxiv=\iftrue     % ARXIV VERSION
\pdfoutput=1

\ifarxiv

\documentclass[12pt,a4paper]{article}
\usepackage[a4paper,text={450pt,650pt},centering]{geometry}

\fi

\ifarxiv\else

\documentclass[11pt,a4paper]{article}
\usepackage{mathptmx}
\usepackage[a4paper,text={130mm,198mm}]{geometry}

\fi

\usepackage{amsmath,amssymb}
\usepackage[bookmarks=true,hyperfigures=true]{hyperref}
\usepackage{graphicx}
\usepackage[nosort]{cite}
\usepackage[bulletsep]{collref}
\let\oldbfseries=\bfseries
\let\oldmdseries=\mdseries
\let\oldnormalfont=\normalfont
\renewcommand{\bfseries}{\oldbfseries\boldmath}
\renewcommand{\mdseries}{\oldmdseries\unboldmath}
\renewcommand{\normalfont}{\oldnormalfont\unboldmath}

\allowdisplaybreaks[3]

\numberwithin{equation}{section}

\usepackage[font=small,labelfont=bf,width=0.85\textwidth]{caption}

\providecommand{\hypersetup}[1]{}
\providecommand{\texorpdfstring}[2]{#1}

\hypersetup{plainpages=false}
\hypersetup{pdfpagemode=UseNone}
\hypersetup{bookmarksnumbered=true}
\hypersetup{pdfstartview=FitH}
\hypersetup{colorlinks=false}
\hypersetup{citebordercolor={.5 1 .5}}
\hypersetup{urlbordercolor={.5 1 1}}
\hypersetup{linkbordercolor={1 .7 .7}}
\providecommand{\href}[2]{#2}
\providecommand{\arxivlink}[1]{\href{http://arxiv.org/abs/#1}{arxiv:#1}}

\newcommand{\N}{\mathcal{N}}
\newcommand{\psu}{\mathfrak{psu}(2,2|4)}

\newcommand{\Tr}{{\rm tr \,}}
\newcommand{\Op}{\mathcal{O}}
\newcommand{\fldZ}{\mathcal{Z}}
\newcommand{\fldX}{\mathcal{X}}
\newcommand{\fldY}{\mathcal{Y}}
\newcommand{\fldD}{\mathcal{D}}
\newcommand{\fldU}{\mathcal{U}}

\newcommand{\fldV}{\mathcal{V}}

\newcommand{\beq}{\begin{equation}}
\newcommand{\eeq}{\end{equation}}

\newcommand{\beqa}{\begin{eqnarray}}
\newcommand{\eeqa}{\end{eqnarray}}

\newcommand{\PTerm}[1]{\{#1\}}
\newcommand{\curlybracket}[2]{\big\{{\textstyle\genfrac{}{}{0pt}{}{#1}{#2}}\big\}}

\newcommand{\nn}{\nonumber}

\begin{document}

%%%%%%%%%%%%%%%%%%%%%%%%%%%%%%%%%%%%%%%%%%%%%%%%%%%%%%%%%%%%%%%%%%%%%%%%%%%%%%%%
%%%%%%%%%%%%%%%%%%%%%%%%%%%%%%%%%%%%%%%%%%%%%%%%%%%%%%%%%%%%%%%%%%%%%%%%%%%%%%%%
% TITLE PAGE

\thispagestyle{empty}
\phantomsection
\addcontentsline{toc}{section}{Title}

\begin{flushright}\footnotesize%
\texttt{IMPERIAL-TP-AR-2010-2},
\texttt{\arxivlink{1012.3985}}\\
overview article: \texttt{\arxivlink{1012.3982}}%
\vspace{1em}%
\end{flushright}

\begingroup\parindent0pt
\begingroup\bfseries\ifarxiv\Large\else\LARGE\fi
\hypersetup{pdftitle={Review of AdS/CFT Integrability, Chapter I.3: Long-range spin chains}}%
Review of AdS/CFT Integrability, Chapter I.3:\\
Long-range spin chains
\par\endgroup
\vspace{1.5em}
\begingroup\ifarxiv\scshape\else\large\fi%
\hypersetup{pdfauthor={Adam Rej}}%
Adam Rej
\par\endgroup
\vspace{1em}
\begingroup\itshape
Blackett Laboratory, Imperial College, London SW7 2AZ, U.K.
\par\endgroup
\vspace{1em}
\begingroup\ttfamily
a.rej@imperial.ac.uk
\par\endgroup
\vspace{1.0em}
\endgroup

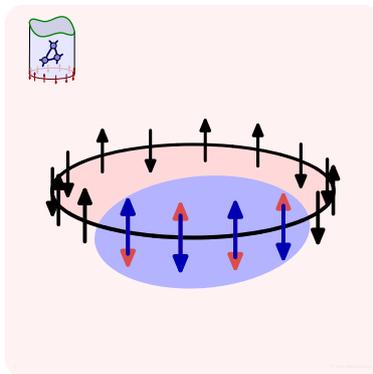
\begin{center}
\includegraphics[width=5cm]{TitleI3.mps}%figure for your chapter
\vspace{1.0em}
\end{center}

\paragraph{Abstract:}
In this contribution we review long-range integrable spin chains that originate from the recently discovered integrability in the planar AdS/CFT correspondence. We also briefly summarise the theory of generic integrable perturbatively long-range spin chains.

\ifarxiv\else
\paragraph{Mathematics Subject Classification (2010):} 
...
 http://www.ams.org/msc
\fi
\hypersetup{pdfsubject={MSC (2010):81T13, 81T60 }}%

\ifarxiv\else
\fi
\hypersetup{pdfkeywords={Long-range spin chains, higher loop corrections, Hubbard model, Inozemtsev model}}%

\newpage

%%%%%%%%%%%%%%%%%%%%%%%%%%%%%%%%%%%%%%%%%%%%%%%%%%%%%%%%%%%%%%%%%%%%%%%%%%%%%%%%
%%%%%%%%%%%%%%%%%%%%%%%%%%%%%%%%%%%%%%%%%%%%%%%%%%%%%%%%%%%%%%%%%%%%%%%%%%%%%%%%
% BODY

%%%%%%%%%%%%%%%%%%%%%%%%%%%%%%%%%%%%%%%%%%%%%%%%%%%%%%%%%%%%%%%%%%%%%%%%%%%%%%%%
\section{Introduction}

The appearance of integrability in the planar AdS/CFT \cite{Maldacena:1997re} is a rather unexpected occurrence. The unravelling of the integrable structures on the gauge theory side of the duality began with the ground-breaking work \cite{Minahan:2002ve}, where the one-loop dilatation operator in the $\mathfrak{so}(6)$ sub-sector has been derived and identified with the Hamiltonian of an integrable $\mathfrak{so}(6)$ spin chain. This was subsequently generalised to the full interaction sector of the theory $\psu$ in \cite{Beisert:2003jj}. At one-loop order the dilatation operator is of the nearest-neighbour type and thus resembles Hamiltonians of other integrable spin chains. At higher orders in perturbation theory, however, this is not the case anymore. The first higher-loop corrections to the dilatation operator were first studied in the $\mathfrak{su}(2)$ sub-sector, see \cite{Beisert:2003tq}, and the two-loop correction found therein has been shown to be integrable as well. Conjecturing the integrability to hold at higher loops and with help of further assumptions, also the three- and four-loop corrections have been found\footnote{The four-loop contribution was only determined up to a single coefficient, which was then uniquely fixed in \cite{Beisert:2003jb}.}. This has furnished first evidence that the integrability might be an all-loop feature of the dilatation operator of $\N=4$ SYM theory. The higher-rank sectors were first studied in \cite{Beisert:2003ys}, where the two- and three-loop corrections to the dilatation operator in the maximal compact sub-sector of the theory $\mathfrak{su}(2|3)$ have been determined and their integrability has been verified. The generalisation to the full theory has turned out to be very intricate, nevertheless higher corrections for the non-compact $\mathfrak{su}(1,1|2)$ sub-sector have been derived in \cite{Zwiebel:2005er} and \cite{Zwiebel:2008gr}. These developments were paralleled by the formulation of the corresponding one-loop and higher-loop Bethe ans\"atze, as well as a host of discoveries of integrable structures on the string theory side. 
Integrable structures have also been found in the context of the $AdS_4/CFT_3$ and $AdS_3/CFT_2$ correspondences. Please refer to other reviews of this series for further details and references. 

The perturbative corrections to the dilatation operator have been found assuming that wrapping interactions may be neglected. These interactions wrap around the chain  and thus account for highly non-local interactions between the spins. Since an interaction between two neighbouring spins contributes a factor $\Op(\lambda)$, first wrapping interactions may in general appear at the order $\Op(\lambda^{L})$, where $L$ is the length of the system. Please refer to \cite{chapHigher} for further discussion of these non-local interactions. In what follows we will always assume that the order of perturbation theory $\ell$ is smaller then the length of the system, i.e. $\ell < L$.

The higher-loop corrections to the dilatation operator exhibit \textit{novel} features when compared with Hamiltonians of the vast majority of integrable spin chains. Firstly, the range of the interactions increases with the loop order. Secondly, beyond the one-loop level operators with the same classical dimension but different lengths are mixed together. The simplest example of such process furnishes the mixing of three scalar fields with two fermions
\beq \label{Lchange}
\Tr \big(\dots \underbrace{\fldX \fldY \fldZ}_{\Delta_0=3, \ L=3} \dots \big) \leftrightarrow \Tr \big(\dots \underbrace{\fldU\fldV}_{\Delta_0=3, \ L=2} \dots \big)\,.
\eeq
Integrable long-range spin chains with these properties have not been hitherto investigated. They should distinguished from the long-range spin chains considered before in the literature, as they are defined as long-range deformations of nearest-neighbour models.  There is a host of evidence that these unusual features do not hinder the integrability. This suggests that integrable perturbatively long-range spin chains should be well-defined and could constitute an interesting class of models not studied in the literature. The Inozemtsev model, see \cite{Inozemtsev:1989yq},  an important intrinsically long-range spin chain and its connection to perturbatively long-range spin chains will be briefly discussed in section \ref{sec: Inozemtsev}. Unless stated otherwise, throughout this review by long-range spin chains we will mean perturbative long-range spin chains. 

The investigation of generic closed integrable long-range spin chains has been initiated in \cite{Beisert:2005wv}, where the underlying symmetry algebra was assumed to be $\mathfrak{gl}(n)$. It has been found that integrable long-range spin chains are characterised by four infinite families of parameters and thus span a very large class. However, it turns out that only two families of the parameters influence the Bethe equations. The two other correspond to rotations of the higher conserved charges and to similarity transformations. The latter do not influence the spectrum. These findings were subsequently generalised  to arbitrary Lie (super)algebra in \cite{Bargheer:2008jt}. Moreover, a novel recursion relation has been proposed, which allows to lift an integrable nearest-neighbour spin chain to its long-range counterpart, see also \cite{Bargheer:2009xy}. This has laid solid foundations for the theory of perturbative long-range systems.

This review is structured as follows. In section \ref{sec:su2} we will briefly discuss the perturbative corrections to the dilatation operator in the $\mathfrak{su}(2)$ sub-sector of the planar $\N=4$ gauge theory. The higher-rank sectors $\mathfrak{su}(2|3)$ and $\mathfrak{su}(1,1|2)$ are the subject of section \ref{sec:highrank}. In section \ref{sec:genspin} we will review the general theory of perturbative long-range integrable spin chains. Finally, in section \ref{sec:hubbard} we will explain an interesting relation between the Hubbard model and long-range spin chains.  In this article we assume that the reader is familiar with the rudiments of integrable spin chains and their application to AdS/CFT correspondence presented in \cite{chapChain}.

\section{The \texorpdfstring{$\mathfrak{su}(2)$}{su(2)} sub-sector} \label{sec:su2}
The $\mathfrak{su}(2)$ sector is one of the simplest dynamical sectors. It has been proven in \cite{Beisert:2003tq} that this sector is closed, i.e. there is no mixing with other types of the operators. It consists of two types of scalar $\fldX$ and $\fldZ$
\beq \label{su2fields0}
\Tr \big(\fldX^M \fldZ^{L-M} \big)+\dots \,.
\eeq
In the spin chain picture one identifies the $\fldX$ fields with say up spins $\uparrow$ and the $\fldZ$ fields with down spins $\downarrow$
\beq \label{su2fields}
\Tr \big(\fldX^M \fldZ^{L-M} \big)+\dots \ \longleftrightarrow \ |\underbrace{\uparrow \uparrow \dots \uparrow}_{M} \underbrace{\downarrow \downarrow \dots \downarrow}_{L-M} \rangle + \dots \,.
\eeq
The cyclicity of the trace imposes closed periodic boundary conditions on the spin chain. Up to now this is merely a change in the notation. The advantage of the spin chain reinterpretation becomes apparent when one considers the one-loop dilatation operator in this sector, which may be extracted from the one-loop $\mathfrak{so}(6)$ dilatation operator found in \cite{Minahan:2002ve} by restricting to the case of two scalar fields. Introducing the notation
\beq \label{perms}
\{n_1,n_2, \ldots, n_l\} = \sum^L_{k=1} P_{k+n_1,k+n_1+1} P_{k+n_2,k+n_2+1} \ldots P_{k+n_l,k+n_l+1}\,,
\eeq
where $P_{a,b}$ permutes the spins at site $a$ and $ b$ in the chain, the one-loop dilatation operator may be written as
\beq \label{su2D2}
D_2 = 2 ( \{\} - \{0\})\,.
\eeq 
Thus $D_2$ is proportional to the Hamiltonian of the $XXX$ spin chain! The computation of higher-loop corrections with diagrammatic methods becomes very involved beyond the leading order. A novel method of determining the higher-loop corrections has been introduced in \cite{Beisert:2003tq}. The authors have analysed and classified the two-loop Green functions corresponding to the operators \eqref{su2fields}. They have advocated that only certain types of interactions are permited, which in the spin chain picture correspond to permutations of the neighbouring sites. Furthermore, it has been argued that at two-loop order only interactions permuting at most three consecutive spins are allowed. One can thus assume that a subclass of \eqref{perms} consisting of all permutations of at most three nearest-neighbours span the basis for the two-loop dilatation operators $D_4$. The coefficients of the linear combinations may be fixed using additional constraints. The simplest one follows from the fact that the scaling dimension of the half-BPS operators $\Tr \fldZ^L$ is protected and does not receive any radiative corrections. Consequently, 
\beq
D_4 \left(\Tr \fldZ^L \right)=0\,,
\eeq
for any $L$. Further constraints follow from the so-called BMN scaling. It has been argued in \cite{Berenstein:2002jq} that the $\ell$-loop anomalous dimension of the operators $\Tr \fldX^M \fldZ^J$ should scale as
\beq
\gamma_{2\ell} \sim (\lambda')^\ell (1+ \Op(1/J))\,, \qquad \lambda'=\frac{g^2}{J^2}\,,
\eeq
for $M=\textrm{fixed}$ and $J\to \infty$. Moreover, the leading coefficient should match the string theory prediction
\beq
\Delta = J +\sum^M_{k=1} \sqrt{1+ 4\pi \lambda' n_k^2}\,.
\eeq
The mode numbers $n_k$ are subjected to the level matching condition $\sum^M_{k=1} n_k =0$. While it is now known that BMN scaling \textit{breaks down} at the four-loop order, see the discussion in \cite{Eden:2006rx}, it has played a major role in the development of the subject. At the two-loop order these both requirement uniquely fix $D_4$ to
\beq \label{su2D4}
D_4 = 2(-4\{\}+6\{0\}-(\{0,1\}+\{1,0\}))\,.
\eeq  
One of the very few manifestations of the integrability at the level of the spectrum are the so-called parity pairs, i.e. pairs of operators with opposite parity and equal energies. Please see review by Charlotte Kristjansen \cite{chapObserv} for the definition of parity and further discussion of parity pairs.  The existence of such pairs hints at the presence of higher conserved charges which commute with the dilatation operator, but anticommute with the parity operator. At one-loop order the simplest of these charges is
\beq
Q^{(2)}_3= 4 (\{1,0\}-\{0,1\})\,.
\eeq
It should be stressed that it is rather a non-trivial task to find explicitly the higher conserved charges for an integrable spin chain. The situation is facilitated to a great extent if the so-called boost operator is known, see \cite{Tetelman} and \cite{WadatiSogo}. Interestingly, as argued in \cite{Grabowski:1994rb}, the mere existence of $Q^{(3)}$ seems to guarantee the existence of all higher charges.

The authors of \cite{Beisert:2003tq} have discovered that the first higher charge may also be determined at the two-loop order such that $[D(\lambda),Q_3 (\lambda)]=0$ holds up to $\Op(\lambda^3)$, i.e.
\beq
\left[D_4, Q^{(2)}_3 \right]+\left[D_2, Q^{(4)}_3 \right]=0\,.
\eeq
This guarantees the degeneracy of the spectrum at two-loop order. It is thus plausible to assume that integrability will be present at higher loops. More generally, if the higher charges are determined to a given loop order $\ell$ and commute with each other up to $\Op(\lambda^{\ell+1})$,  the system is said to be perturbatively integrable up to $\ell$-th order. 

There is strong evidence that the $\mathfrak{su}(2)$ sector is perturbatively integrable at least up to three-loop order. The three-loop dilatation operator may be again found \cite{Beisert:2003tq} by imposing the degeneracy for the paired operators (i.e. imposing the presence of the parity pairs) in conjunction with the constraints discussed above 
\beq \label{su2D6}
D_6=4\big(15\{\}-26\{0\}+6(\{0,1\}+\{1,0\})+\{0,2\}-(\{0,1,2\}+\{2,1,0\})\big)\,.
\eeq
Also the corresponding three-loop correction to the first higher charge satisfying the perturbative integrability condition at three-loop order may be found. The same set of conditions allowed to constrain the form of the four-loop correction to the dilatation operator up to two coefficients \cite{Beisert:2003tq}. Moreover, it has been found that one of these unknowns does not affect the spectrum since it can be eliminated by a similarity transformation
\beq \label{similarity}
D' = J(\lambda)\, D \,J(\lambda)^{-1}\,.
\eeq
In \cite{Beisert:2003jb} the remaining constant has been fixed by a more careful analysis of the implications of the BMN limit. This analysis has been further extended to the five-loop order in \cite{Beisert:2004hm}. In \cite{Ryzhov:2004nz} it has been argued that the BMN limit is sufficient to determine the all-loop two-spin interaction part of the dilatation operator.
One should however note that it is \textit{incorrect} to assume the BMN limit at and beyond four-loop order and the corrections found with help of this constraint need to be modified. It has been proposed in \cite{Beisert:2007hz} to use instead the form of the one-magnon dispersion relation together with the two-magnon scattering matrix derived in \cite{Beisert:2005tm}. This allowed to determine the  four-loop correction up to an unknown constant $\beta^{(4)}_{2,3}$ and parameters related to the similarity transformations, \textit{cf.} \eqref{similarity}. It turns out that the constant $\beta^{(4)}_{2,3}$ multiplies a term with four permutations that reshuffle only four consecutive spins and thus may be determined by evaluating only a sub-class of the Feynman diagrams. These diagrams have been calculated in \cite{Beisert:2007hz} and the remaining coefficient could have been fixed to $\beta^{(4)}_{2,3}=4\zeta(3)$. This is the first evidence of the so-called dressing phase introduced in section \ref{sec:genspin}. For a discussion of the dressing factor of the AdS/CFT correspondence the reader should refer to the review by Pedro Vieira and Dmytro Volin \cite{chapSProp}.
\section{Higher-rank sectors : \texorpdfstring{$\mathfrak{su}(2|3)$}{su(2|3)} and \texorpdfstring{$\mathfrak{su}(1,1|2)$}{su(1,1|2)}} \label{sec:highrank}
In this section we will discuss higher-order corrections to the dilatation operator beyond the $\mathfrak{su}(2)$ sub-sector. The novel feature, when compared with the previous case, is the central role played by the symmetry algebra. The higher-loop corrections to the symmetry generators are strongly constrained by the algebra relations
\beq \label{algebrarel}
[J^{A}(\lambda), J^{B}(\lambda)]=f^{A B}_{C} J^C(\lambda)\,.
\eeq
The structure constants $f^{AB}_C$ \textit{do not} receive quantum corrections. In what follows we will discuss two particular examples: $\mathfrak{su}(2|3)$ and $\mathfrak{su}(1,1|2)$ sub-sectors.
\subsection{The maximal compact sub-sector \texorpdfstring{$\mathfrak{su}(2|3)$}{su(2|3)}}
The $\mathfrak{su}(2|3)$ sector consists of three scalars and two fermionic fields and can be schematically represented by
\beq \label{fieldcontent1}
\Tr \big( \fldX^{M_1} \fldY^{M_2}  \fldU^{M_3} \fldV^{M_4} \fldZ^{L-M} \big)+\ldots \,,
\eeq
where $M=M_1+M_2+M_3+M_4$. Please note that in view of the mixing processes \eqref{Lchange} the length $L$ is not conserved beyond the one-loop order. A generic state of the $\N=4$ SYM theory is characterised by the classical dimension $\Delta_0$, the $\mathfrak{su}(2)^2$ labels $[s_1, s_2]$, the $\mathfrak{su}(4)$ Dynkin labels $[q_1,p,q_2]$, the $\mathfrak{u}(1)$ hypercharge $B$ and the length $L$. Please refer to \cite {chapChain} for details. The truncation to the $\mathfrak{su}(2|3)$ sector is obtained by restricting to the states with 
\beq
\Delta_0 = p+\tfrac{1}{2}q _1+ \tfrac{3}{2} q_2 \,.
\eeq
This also implies certain relations on some of the generators, see \cite{Beisert:2003ys}. The full symmetry algebra $\psu$ thus effectively reduces to $\mathfrak{su}(2|3)$. It consists of the generators
\beq
J=\{ L^{\alpha}_{\phantom{\alpha} \beta}, R^a_{\phantom{a}b}, D, \delta D \,|\, Q^a_{\phantom{\alpha} \alpha}, S^{\alpha}_{\phantom{a}a} \}\,.
\eeq
The $\mathfrak{su}(2)$ and $\mathfrak{su}(3)$ generators $L^{\alpha}_{\phantom{\alpha}\beta}$ and $R^a_{\phantom{a}b}$ are traceless. The corresponding commutation relations are as follows
\beqa \label{relb}
&&[L^{\alpha}_{\phantom{\alpha} \beta}, \ J_{\gamma}]=\delta^{\alpha}_{\gamma} J_{\beta}-\tfrac{1}{2} \delta^{\alpha}_{\beta} J_{\gamma}\,, \quad [L^{\alpha}_{\phantom{\alpha} \beta}, \ J^{\gamma}]= -\delta^{\gamma}_{\beta} J^{\alpha} + \tfrac{1}{2} \delta^{\alpha}_{\beta} J^{\gamma}\,, \\
&&[R^a_{\phantom{a}b}, \ J_c]=\delta^a_c J_b - \tfrac{1}{3} \delta^{a}_b J_c\,, \quad [R^a_{\phantom{a}b}, \ J^c]=-\delta^c_b J^a+\tfrac{1}{3} \delta^a_b J^c\,.
\eeqa
The commutators of the dilatation operator and its anomalous part are given by
\beq \label{DdeltaD}
[D, \ J]= \text{eng}(J) J\,, \qquad [\delta D, \ J]=0\,,
\eeq
with $\text{eng}(Q)=-\text{eng}(S)=\frac{1}{2}$. The supercharges $Q^a_{\phantom{\alpha} \alpha}$ and $S^{\alpha}_{\phantom{a}a}$ anticommute \footnote{The supersymmetry generator $Q^a_{\phantom{a} \alpha}$ should not be confused with the higher conserved charges $Q_r$. Even though the same symbol is used to denote both charges, it will become clear from the context which quantity is referred to.} 
\beq \label{rele}
\{S^{\alpha}_{\phantom{a}a}, \ Q^b_{\phantom{\alpha} \beta}\}=\delta^b_a L^{\alpha}_{\phantom{\beta}\beta}+\delta^{\alpha}_{\beta} R^b_{\phantom{a} a}+\tfrac{1}{6}\delta^b_a \delta^{\alpha}_{\beta} \big(2\, D +\delta D \big)\,.
\eeq
The symmetry generators act on \eqref{fieldcontent1} by reshuffling the  operators in the trace and changing the labels $M_1, M_2, M_3, M_4$ and $L$. An interaction replacing the sequence of fields $A_1, \dots A_n$  within the state $|C_1 \dots C_L \rangle = (-1)^{(C_1 \dots C_i)(C_{i+1}\dots C_L)} |C_{i+1} \dots C_L C_1 \dots C_{i} \rangle$ by $B_1, \dots, B_m$ will be denoted as
\beqa \nonumber
&&\curlybracket{ A_1 \dots A_n }{ B_1 \dots B_m }  |C_1 \dots C_L \rangle= \\
&&\sum^{L-1}_{i=0} (-1)^{(C_1 \dots C_i)(C_{i+1}\dots C_L)}\delta^{A_1}_{C_{i+1}} \dots \delta^{A_n}_{C_{i+n}} | B_1 \dots B_m C_{i+n+1} \dots C_L C_1 \dots C_i \rangle\,.
\eeqa
Here $(-1)^{XY}$ equals $-1$ if both $X$ and $Y$ are fermionic and $+1$ otherwise.

The key observation of \cite{Beisert:2003ys} is that the algebra relation \eqref{relb}-\eqref{rele} \textit{largely} constrain the form of the generators. For example, at tree-level one expects the following general $\mathfrak{su}(3)\times \mathfrak{su}(2)$ invariant form of the generators
\beqa \label{R}
R^{a}_{\phantom{b}b}&=&c_1\, \curlybracket{a}{b}+c_2 \, \delta^a_b \curlybracket{c}{c}\,,\\ \label{L}
L^{\alpha}_{\phantom{\beta}\beta}&=&c_3 \, \curlybracket{ \alpha}{\beta}+c_4 \, \delta^{\alpha}_{\beta} \curlybracket{ \gamma}{\gamma}\,, \\ 
D_0&=&c_5 \, \curlybracket{a}{a}+c_6 \,\curlybracket{\alpha}{\alpha}\,,\\
(Q_0)^a_{\alpha}&=&c_7 \, \curlybracket{a}{\alpha}\,, \\
(S_0)^{\alpha}_{a}&=&c_8 \,\curlybracket{\alpha}{a}
\,.
\eeqa
Please note that the generators $R^a_{\phantom{b}b}$ and $L^{\alpha}_{\phantom{\beta}\beta}$ are not influenced by radiative corrections and the formulas \eqref{R} and \eqref{L} will be thus valid to all orders. The non-trivial solution to \eqref{relb}-\eqref{rele} is furnished by
\beq
c_1=c_3=c_5=1\,, \quad c_2=-\tfrac{1}{3}\,, \quad c_4=-\tfrac{1}{2}\,,\quad c_6=\tfrac{3}{2}\,,\quad c_7=e^{i \beta}\,, \quad  c_8=e^{-i \beta}\,.
\eeq
Moreover, the parameter $\beta$ corresponds to the similarity transformation
\beq
J_0 \to e^{2 \,i \,\beta \,D_0}\, J_0 \, e^{-2\, i \,\beta \,D_0}\,.
\eeq
Thus, the commutation relations allowed to unambiguously determine the form of the generators! A similar method has been applied in \cite{Beisert:2003ys} to determine corrections to the generators $Q$ and $S$ up to the order $\Op(\lambda^2)$ and up to the order $\Op(\lambda^3)$ for the dilatation generator $D$. Please note that since the perturbative expansion of $\delta D$ starts at $\Op(\lambda)$ and in view of \eqref{DdeltaD} the $k$-th order contribution to $\delta D$ may be constrained through the perturbative expansion
of the remaining generators up to the order $\Op(\lambda^{(k-1)})$. At higher orders, however, the relations \eqref{relb}-\eqref{rele} \textit{do not} determine all physical coefficients and further assumptions must be made. Up to the three-loop order it is sufficient to exploit constraints following from the topology of the Feynman diagrams together with the absence of the radiative corrections for the half-BPS states and impose the BMN limit, see \cite{Beisert:2003ys}. The two- and three-loop corrections to the dilatation operator found in this way preserve the maximum amount of parity pairs and the dilatation operator was conjectured to be perturbatively integrable up to three-loop order \cite{Beisert:2003ys}. The next conserved charge $Q_3$ has been constructed in \cite{Agarwal:2005jj} up to the order $\Op(\lambda^2)$.

In \cite{Beisert:2008qy} it has been proposed how to reformulate the description of the $\mathfrak{su}(2|3)$ spin-chain in order to eliminate the length-changing processes \eqref{Lchange}. The underlying idea is to ``freeze out'' the dynamic effects by choosing one of the bosonic fields, say $\phi^3:=\fldZ$ as the background field. The other fields in the sector are then redefined as follows
\beq
\{\phi^1,\ \phi^2, \ \psi^1,\ \psi^2 \}\ni \mathcal{F} \mapsto \mathcal{F}_n :=\mathcal{F} \underbrace{\fldZ \dots \fldZ}_{n} \,.
\eeq
In this way the dynamic effects are traded for infinitely many spin degrees of freedom labelled by $n$ and the spin chain becomes static. This reformulation may be useful to make the dynamic spin chains accessible to an algebraic treatment.
\subsection{The non-compact \texorpdfstring{$\mathfrak{su}(1,1|2)$}{su(1,1|2)} sub-sector}
The constraints following from algebra relations become particularly important in the non-compact sectors, where the modules are infinite-dimensional. Any diagrammatic calculations in this case are only realistic at low loop order, as for example at the two-loop level in the fermionic $\mathfrak{sl}(2)$ sub-sector \cite{Belitsky:2005bu}. The algebraic approach in non-compact sectors has been advocated in \cite{Zwiebel:2005er} and the complete $\Op(\lambda^{3/2})$ symmetry algebra in the $\mathfrak{su}(1,1|2)$ sub-sector as well as the two-loop correction to the dilatation operator have been found.
The $\mathfrak{su}(1,1|2)$ sub-sector consists of two scalar fields, two fermions and derivatives
\beq
\fldD^k \fldZ\,, \quad \fldD^k \fldX\,, \quad \fldD^k \fldU, \quad \fldD^k \dot{\fldU} \,.
\eeq
Formally, the truncation of the full symmetry algebra to the $\mathfrak{su}(1,1|2)$ sub-sector is achieved by setting the classical dimensions of states simultaneously equal to the following linear combination of the eigenvalues of the Cartan generators of the $\psu$ algebra
\beq
D_0 = s_1 +\tfrac{1}{2}q_2+p+\tfrac{3}{2} q_1 = s_2+\tfrac{1}{2} q_1 +p +\tfrac{3}{2} q_2\,.
\eeq
Interestingly, the residual symmetry is larger than expected and consists of a tensor product $\mathfrak{psu}(1,1|2) \times \left(\mathfrak{psu}(1|1)\right)^2$. The anomalous part of the dilatation operator $\delta D$ is a central charge for both components of the product. The full set of commutation relations may be found in \cite{Zwiebel:2005er}.

By invoking constraints from Feynman rules, imposing the algebra relations \eqref{algebrarel} and using representation theory, it has been found in \cite{Zwiebel:2005er} that the next-to-leading corrections\footnote{The generators of $\mathfrak{psu}(1,1|2)$ have an expansion in $g^2 \sim \lambda$, while the expansion parameter of the $\mathfrak{psu}(1|1)$ generators is $g \sim \sqrt{\lambda}$.} to the $\mathfrak{psu}(1,1|2)$ algebra generators satisfy
\beq \label{JNLOJLO}
J_{\text{NLO}}=\pm \ [J_{\text{LO}}, \ X]\,.
\eeq
The sign in front of the commutator is different for generators corresponding to positive and negative algebra roots. The generator $X$ may be expressed through the $\mathfrak{psu}(1|1)^2$ supercharges $T^{\pm}$ and $\bar{T}^{\pm}$ together with an auxiliary generator $h$
\beq \label{XLO}
X=\tfrac{1}{2} \left( \{\bar{T}^-,[\bar{T}^{+}, h]\}-\{T^{+},[T^{-},h]\} \right)\,.
\eeq
The generator $h$ at the leading order is a one-site generator of the harmonic numbers $H(j)$
\beqa \nonumber
h \, |\fldD^k \fldZ \rangle &=&H(k)\,|\fldD^k \fldZ\rangle\,,\quad h \,|\fldD^k \fldX\rangle = H(k)\,| \fldD^k \fldX \rangle \\
h \, |\fldD^k \fldU \rangle &=&H(k+1)\,|\fldD^k \fldU \rangle\,,\quad h \,|\fldD^k \dot{\fldU}\rangle = H(k+1)\,| \fldD^k \dot{\fldU} \rangle \,.
\eeqa
The higher corrections to $h$ may be found recursively \cite{Zwiebel:2005er}. Also the $\Op(\lambda^{3/2})$ corrections to the fermionic generators of the two copies of $\mathfrak{psu}(1|1)$, that is $T^{\pm}$ and $\bar{T}^{\pm}$, could have been determined in a compact form. Since the classical action of these generators is trivial, this is enough to determine the two-loop dilatation generator
\beq
\delta D^{\mathfrak{su}(1,1|2)}_4 =2 \left\{\bar{T}^+, \bar{T}^-\right\}_4= 2 \left\{T^+, T^-\right\}_4= 2 \left\{T^+_3, T^-_1\right\}+ 2 \left\{T^+_1, T^-_3\right\}\,.
\eeq
The two-loop correction determined in this way was found to reproduce correctly the two-loop anomalous dimension in the $\mathfrak{sl}(2)$ and $\mathfrak{su}(1|1)$ sub-sectors, at least for the states considered \cite{Zwiebel:2005er}. It has been argued in \cite{Zwiebel:2008gr} that the relation \eqref{JNLOJLO} has a very simple generalisation at higher orders
\beq \label{DJJ}
\frac{\partial}{\partial \lambda} J(\lambda) = \pm \ [J(\lambda), \ X(\lambda)]\,.
\eeq
In other words, $X(\lambda)$ generates translations in $\lambda$ for symmetry generators. The leading order result \eqref{XLO} is lifted to higher orders in the simplest possible way
\beq
X(\lambda)=\tfrac{1}{2} \left( \{\bar{T}^-(\lambda),[\bar{T}^{+}(\lambda), h(\lambda)]\}-\{T^{+}(\lambda),[T^{-}(\lambda),h(\lambda)]\} \right)\,.
\eeq
The function $h(\lambda)$ may be recursively determined from the corresponding Serre-like relations, see \cite{Zwiebel:2008gr} for further details. The equation \eqref{DJJ} allowed to determine the dilatation operator in this sector up to three-loop order, which was subsequently subject to numerous spectral tests (see \cite{Zwiebel:2005er} and \cite{Zwiebel:2008gr}) and appears to be perturbatively integrable.

\section{Generic integrable long-range spin chains}\label{sec:genspin}
The $\mathfrak{su}(2)$, $\mathfrak{su}(2|3)$ and $\mathfrak{su}(1,1|2)$ spin chains discussed above furnish examples of novel long-range integrable spin chains. The integrability of any spin chain is based on the existence of an infinite set of independent hermitian commuting charges $Q_r$
\beq \label{chargescommute}
[Q_r,\ Q_s]=0\,.
\eeq
The $Q_2$ charge is usually associated with the Hamiltonian, while the total momentum operator is usually identified with $\text{exp}( i \,Q_1)$. It is an interesting question what are the generic long-range spin chains satisfying \eqref{chargescommute}. In this section we will discuss the recent progress in the theory of such systems. 

In this section we will assume that the spin chain charges admit perturbative expansion
\beq  \label{Qperturbative}
Q_r(\lambda) = \sum^{\infty}_{k=0} \left(\tfrac{\lambda}{16 \,\pi^2}\right)^{k}\, Q^{(k)}_r\,.
\eeq
Furthermore, we will assume that the maximal range of $Q^{(k)}_r$ is $r+k$, i.e. $Q^{(k)}_r$ acts locally on $r+k$ adjacent sites in the spin chain. Please note that for finite values of $\lambda$ the range of interactions becomes formally infinite.
\subsection{Closed long-range spin chains with \texorpdfstring{$\mathfrak{gl}(n)$}{gl(n)} symmetry algebra} \label{sec:gln}
Generic spin chains with the underlying symmetry algebra $\mathfrak{gl}(n)$ have been investigated in \cite{Beisert:2005wv}. It has been proposed that the $\mathfrak{gl}(n)$-invariant long-range interactions may be expanded in the basis \eqref{perms}. The range of an interaction $\{n_1,\dots, n_l\}$ is given by $R=\text{max}\{n_i\}-\text{min}\{n_i\}+2$. Consequently, the basis for the $k$-loop correction to the charge $Q_r$ is spanned by \eqref{perms} with $R \leq r+k$. The number of all permutations up to range $R$ is given by $R!-(R-1)!+1$. Please note that at the $k$-loop order the relation \eqref{chargescommute} amounts to
\beq \label{chargescommutekorder}
\sum^k_{j=0} \, [Q^{(j)}_r,\ Q^{(k-j)}_s]=0\,,
\eeq
so that the procedure is recursive. The authors of \cite{Beisert:2005wv} have applied this method to $Q_2$ and $Q_3$ charges up to and including four-loop order. Interestingly, it is enough to consider solely commutation relations between $Q_2$ and $Q_3$ since the commutators with higher charges \textit{do not} lead to further restrictions. The relation \eqref{chargescommutekorder} does not fix all the coefficients of the basis. For example, the $Q_2$ charge up to two-loop order is presented in Table \ref{tab:Q2}. The free parameters appearing at any loop order can be divided into three classes, which we will discuss in what follows.

The first class constitute the moduli $\alpha_l(\lambda)$ and $\beta_{r,s}(\lambda)$. They govern propagation and scattering of the spins and differ for different models. They enter directly into the Bethe equations and dispersion relation. It has been conjectured in \cite{Beisert:2005wv} that only the main equation out of the set of Bethe equations corresponding to the nearest-neighbour integrable $\mathfrak{gl}(n)$ spin chain needs to be modified. Explicitly, the main Bethe equations take the following form
\beq \label{glnBE}
1=\left(\frac{x(u_{k}-\tfrac{i}{2})}{x(u_{k}+\tfrac{i}{2})}\right)^L \prod^{K_u}_{j=1, j\neq k}\, \frac{u_{k}-u_{j}+i}{u_{k}-u_{j}-i}\text{exp}\left(2\,i\,\theta(u_{k},u_{j}) \right) \ \prod^{K_v}_{l=1} \,\frac{u_{k}-v_{l}-\tfrac{i}{2}}{u_{k}-v_{l}+\tfrac{i}{2}} \,.
\eeq
The reader might find it useful to refer to \cite{chapABA} and\cite{chapSMat} for a pedagogical discussion of single-level and nested Bethe equations. Here, the main Bethe roots are labelled by $u_k$, while the auxiliary Bethe roots coupling to the main roots are denoted by $v_j$. The difference to the Bethe equations of the nearest-neighbour spin chain is twofold. Firstly, the function $x(u)$, the so-called rapidity map, determines the momentum-rapidity relation of a single magnon
\beq
\text{exp}(i\,p(u)) = \frac{x(u+\tfrac{i}{2})}{x(u-\tfrac{i}{2})}\,.
\eeq
The rapidity map depends on the $\alpha_l(\lambda)$ parameters through the relation
\beq
u(x)=x+\sum^{\infty}_{l=0} \frac{\alpha_l(\lambda)}{x^{l+1}}\,,
\eeq
which needs to be solved for $x$. Secondly, the additional piece of the scattering matrix, 
$\text{exp}(2\,i\,\theta(u,v))$, known in the literature as the dressing factor, is determined by the $\beta_{r,s}(\lambda)$ parameters
\beq
\theta(u,v)=\sum^{\infty}_{r=2} \sum^{\infty}_{s=r+1} \beta_{r,s}(\lambda) \left( q_r (u)\, q_s(v)-q_s(u)\,q_r(v)\right)\,.
\eeq
The $\beta_{r,s}(\lambda)$ coefficients start at order $\Op(\lambda^{s-1})$
\beq
\beta_{r,s}(\lambda)=\sum^{\infty}_{k=s-1}\left(\tfrac{\lambda}{16\,\pi^2}\right)^{k} \beta^{(k)}_{r,s}\,.
\eeq
The parity conservation requires $\beta_{r,s}=0$ for all even $r+s$. The quantities $q_r(u)$ are the elementary magnon charges and are given by
\beq \label{glnqr}
q_r(u)=\frac{i}{r-1}\left(\frac{1}{x(u+\tfrac{i}{2})^{r-1}}-\frac{1}{x(u-\tfrac{i}{2})^{r-1}}\right)\,.
\eeq
Clearly, the distinct character of the $\alpha_l(\lambda)$ and $\beta_{r,s}(\lambda)$ moduli parameters becomes apparent. The $\alpha_l(\lambda)$ parameters specify the one-magnon state, while the $\beta_{r,s}(\lambda)$ ``dress'' the scalar part of the scattering matrix of two magnons. Thanks to integrability these pieces of information are enough to fully describe the system. 

The second class of parameters $\gamma_{r,s}(\lambda)$ are elements of the normalisation matrix of the charges. Upon introducing the normalised charges, for which the eigenvalues are given by a sum over the charge densities $\tilde{Q}_s:=\sum^{K_u}_{k=1} q_s (u_k)$, the $[\gamma(\lambda)]_{r,s}$ matrix simply acts as a rotation matrix
\beq
Q_r = \gamma_{r,0}(\lambda)\,L + \sum^{\infty}_{s=2} \gamma_{r,s}(\lambda) \tilde{Q}_s\,.
\eeq
This transformation readily preserves the commutation relations \eqref{chargescommute}.

Finally, the last class is spanned by the parameters $\epsilon_{k,l}(\lambda)$, which merely influence the eigenvectors and correspond to similarity transformations. They are thus unphysical.

The authors of \cite{Beisert:2005wv} have only analysed $Q_2$ and $Q_3$ charges. Although it seems very plausible that all charges may be constructed in this way, it was still rather a hypothesis. The integrability of the long-range spin chains with the $\mathfrak{gl}(n)$ symmetry algebra has been first confirmed  in \cite{Beisert:2007jv} by constructing the corresponding Yangian algebra up to and including three-loop order. Please refer to \cite{chapYang} for details on Yangians and their relation to integrability.

\subsection{Generic integrable long-range spin chains}
A method for constructing integrable closed long-range spin chains with generic Lie (super)algebras and spin representations has been introduced in \cite{Bargheer:2008jt,Bargheer:2009xy} inspired by the findings of \cite{Beisert:2008cf}. Interestingly, it is a bottom-up approach. The starting point provides an integrable nearest-neighbour spin chain with a symmetry (super)algebra $\mathcal{A}$ and a given spin representation. It has been proposed that the higher-loop deformations of the conserved charges are governed by a generating equation similar to \eqref{DJJ}
\beq \label{DQQ}
\frac{d}{d \lambda} Q_r(\lambda)=i\,[X (\lambda),\ Q_r(\lambda)]+\sum^{\infty}_{s=2}\gamma_{r,s}(\lambda) Q_s(\lambda)\,.
\eeq
Here, $X(\lambda)$ is some operator with well-defined commutation relations with all conserved charges. It is straightforward to check that the deformations generated by \eqref{DQQ} preserve the commutation relations \eqref{chargescommute}. Substituting the expansion \eqref{Qperturbative} into \eqref{DQQ} one can order by order ``boost'' an integrable nearest-neighbour spin chain to its long-range counterpart. The freedom encountered in the previous sub-section while determining the generic form of the higher-loop corrections corresponds to freedom in choosing the $X(\lambda)$ operator. It has been advocated in \cite{Bargheer:2008jt,Bargheer:2009xy} that there are three different admissible classes of such operators: boost charges, bi-local charges and local charges. The first two act inhomogeneously on the spin chain and are parametrised by $\alpha_r(\lambda)$ and $\beta_{r,s}(\lambda)$ respectively. The local operators, on the other hand,  do not influence the spectrum and thus may be associated with the $\epsilon_{k,l}(\lambda)$ degrees of freedom. The Bethe equation diagonalising spin chains constructed in such way are similar to those presented in sub-section \ref{sec:gln}
\beq \label{gscBE}
1=\left(\frac{x(u_{k}-\tfrac{i}{2}\,t_a)}{x(u_{k}+\tfrac{i}{2}\,t_a)}\right)^L \mathop{\prod^{r}_{b=1} \prod^{K_b}_{j=1}}_{(b,j)\neq(a,k)} \frac{u_{a,k}-u_{b,j}+\tfrac{i}{2} C_{ab}}{u_{a,k}-u_{b,j}-\tfrac{i}{2}C_{ab}}\,\text{exp}(2\,i\,\theta^{\{t\}}(u_{a,k},u_{b,j}))\,.
\eeq
The number of levels of the Bethe equations $r$ coincides with the rank of the Lie (super)algebra $\mathcal{A}$. The Dynkin labels of the spin representation are denoted by $t_a,\,a=1,\dots,r$ and the symmetric Cartan matrix is represented by $C_{ab}$. The dressing phase $\theta^{\{t\}}$ is indexed with $t$ to remind that the elementary magnon charges are also influenced by the spin representation
\beq
q_r(t,u)=\frac{i}{r-1} \left(\frac{1}{x(u+\tfrac{i}{2}\,t)}-\frac{1}{x(u-\tfrac{i}{2}\,t)}\right)\,.
\eeq
These results were obtained by applying asymptotic Bethe ansatz techniques to the chain constructed by means of \eqref{DQQ}. Equation \eqref{DQQ} thus plays a central role in the theory of closed long-range integrable spin chains.

In \cite{Beisert:2008cf} the most general perturbatively long-range integrable spin chains in the fundamental representation of the $\mathfrak{gl}(n)$ symmetry algebra and with open boundary conditions have been studied. For open spin chains any excitation returns back to its initial position after being shifted $2L$ times. On its way it is reflected at the two boundaries, each of them giving rise to a boundary scattering phase. Moreover, in general, the momentum after reflection is not equal to reversed incoming momentum and the relation between those two momenta needs to be specified via the reflection map. This is due to the fact that the Hamiltonian will generically not preserve parity. Thus the corresponding Bethe equations differ structurally from the Bethe equations for the closed chains. A set of such Bethe equations for arbitrary boundary scattering phase has been formulated in \cite{Beisert:2008cf}.

\subsection{Examples:  Inozemtsev spin chain} \label{sec: Inozemtsev}
In \cite{Serban:2004jf} the first attempt has been made to embed the novel perturbative long-range integrability in the framework of well-studied integrable models. It was found that up to three-loop order the dilatation operator in the $\mathfrak{su}(2)$ sector may be constructed from the conserved charges of the Inozemtsev model \cite{Inozemtsev:1989yq}. 

The Inozemtsev model furnishes one of the few known examples of integrable long-range spin chains which are not defined as a deformations of nearest-neighbour models. The Hamiltonian of this model is given by
\beq \label{Inozemtsev}
H=\sum^L_{j=1} \sum^{L-1}_{n=1} f_{L, \kappa}(n)(1-P_{j,j+n})\,,
\eeq
where $P_{a,b}$, as before, denotes the permutation of sites $a$ and $b$. The spin chain is assumed to be in the fundamental representation of the $\mathfrak{su}(2)$ symmetry algebra. The interaction strength $f_{L, \kappa}(n)$ is given by the elliptic Weierstrass function
\beq
f_{L, \kappa}(z)=\tfrac{1}{z^2}+\mathop{\sum \hspace{0.1mm}'}^{\infty}_{m,n=-\infty}\left(\frac{1}{(z-m\,L\,-i\,n\,\pi/\kappa)^2}-\frac{1}{(m\,L\,+i\,n\,\pi/\kappa)^2}\right)\,,
\eeq
where the prime means that the term $m=n=0$ should be omitted. A detailed study of the Hamiltonian \eqref{Inozemtsev}, see \cite{Inozemtsev:1989yq},  gave compelling evidence in favour of its integrability. In particular, the corresponding Lax pair has been found. In the limit $\kappa \to 0$ the interaction interpolates smoothly to the Haldane-Shastry interaction \cite{Haldane:1987gg}-\cite{Shastry:1987gh}, which is another known example of an integrable long-range spin chain.

The authors of \cite{Serban:2004jf} have found that simple linear combinations of the higher conserved charges of the Inozemtsev model allow to reconstruct the dilatation operator in the $\mathfrak{su}(2)$ model up to three-loop order. It is necessary to invoke the higher charges since the Hamiltonian \eqref{Inozemtsev} only involves two spin interactions, while already at three-loop order the $\mathfrak{su}(2)$ dilatation operator acts on three sites simultaneously. Under a suitable identification of the coupling constant
\beq
\frac{\lambda}{16\,\pi^2}=\sum_{n>0}\frac{1}{4\, \sinh^2(n\,\kappa)}
\eeq
and keeping $\frac{\lambda}{16 \,\pi^2}$ perturbatively small, the Inozemtsev model turns into long-range model of the type discussed in \ref{sec:gln}. Up to four-loop order
\beq
\alpha_l = \lambda \delta_{l,0}+\lambda^3+\Op(\lambda^4)\,, \qquad \beta_{r,s}=0+\Op(\lambda^4)\,,
\eeq
\beq
\gamma_{2,r}=(2+6\,\lambda-20\,\lambda^2+120\,\lambda^3)\,\delta_{r,2}+(6\,\lambda^2-30\,\lambda^3)\,\delta_{r,4}+\Op(\lambda^4)\,.
\eeq
It would be interesting to find higher-loop corrections to the above formulas.

\section{Hubbard model}\label{sec:hubbard}

In this section we will discuss an intriguing relation between a short-range dynamical model of electrons, the Hubbard model, and the long-range spin chains discussed before. 

The Hubbard model is a dynamical, short-range model of $N$ electrons on $L$ lattice sites. Due to Pauli's exclusion principle, there are four possible states on each lattice site: no particle, spin-up electron, spin down electron and
double occupied state with spin-up and spin-down electrons. In what follows, we will consider the half-filled case $N=L$. The Hamiltonian of the Hubbard model consists of the kinetic part that forces the electrons to jump between different sites and the potential part, which according to the value of $U$ corresponds to repulsive or attractive force
\beq
\label{H}
\hat{H}_{{\rm Hubbard}}=
-t\, \sum_{i=1}^L \sum_{\sigma=\uparrow,\downarrow}
\left(c^\dagger_{i,\sigma} c_{i+1,\sigma}+
c^\dagger_{i+1,\sigma} c_{i,\sigma}\right)+
t\,U\, \sum_{i=1}^L
c^\dagger_{i,\uparrow} c_{i,\uparrow}c^\dagger_{i,\downarrow} c_{i,\downarrow}\, .
\eeq
The operators $c^\dagger_{i,\sigma}$ and $c_{i,\sigma}$ are canonical Fermi operators obeying standard anticommutation relations. We assume the system to be closed and thus we identify
$
c_{L+1,\sigma}=c_{1,\sigma}$ and $c^\dagger _{L+1,\sigma}=c^\dagger _{1,\sigma}$. The Hamiltonian is invariant with respect to the $\mathfrak{su}(2)$ transformations
\beq
[\hat{H}_{\rm Hubbard},\ \hat{S}^{\,a}]=0\,, \qquad \qquad a=+,-,z \,,
\eeq
with $\hat{S}^{\,a}=\sum^{L}_{i=1} \hat{S}^{\,a}_i$. This allows to classify the spectrum according to the eigenvalues of the total spin an its $z$ component. The integrability of this model has been shown in \cite{Lieb:1968zza}.

It has been shown in \cite{Rej:2005qt} that upon the following identification of the parameters
\beq \label{tUident}
t=-\frac{2\,\pi}{\sqrt{\lambda}}\,,
\qquad \qquad
U=\frac{4\,\pi}{\sqrt{\lambda}}\,,
\eeq
this short range model may be identified with the BDS spin chain \cite{Beisert:2004hm}. Please note that with the identification \eqref{tUident}  and in the limit $\lambda \to 0$ the potential part of the Hamiltonian is dominating and  perturbation theory around the states with minimal potential energy may be applied. This allowed to show that the effective Hamiltonian acting on the ground state space of the potential part coincides at one-, two- and three-loop order with the corresponding dilatation operator in the $\mathfrak{su}(2)$ sub-sector, \textit{cf.} formulas \eqref{su2D2}, \eqref{su2D4} and \eqref{su2D6}. The reader should note that the ground space of the potential part of the Hamiltonian \eqref{H} is identical with the Hilbert space of a $\mathfrak{su}(2)$ spin chain. Please refer to \cite{Rej:2005qt} for detailed description of this procedure. Moreover, the spectral equations of the Hubbard model (Lieb-Wu equations \cite{Lieb:1968zza}) have been shown to reproduce to any perturbative order the Bethe equations of the long-range spin chain with the $\mathfrak{su}(2)$ symmetry algebra and the following moduli parameters
\beq
\alpha_l = \lambda \delta_{l,0}\,, \qquad \beta_{r,s}=0\,.
\eeq
Even though this choice \textit{disagrees} with the asymptotic Bethe equations of the $\mathfrak{su}(2)$ sub-sector of $\N=$ SYM at four-loop order and beyond, see \cite{chapSProp}, it may suggest that generic long-range spin chains as well as the asymptotic integrability in the $\N=4$ SYM theory may be intimately related to yet-to-be-discovered integrable short-range models.

\section{Conclusions}
Integrable long-range spin chains are a natural and very non-trivial extension of the nearest-neighbour spin chains, a prime example in the literature on integrable models. The complexity of the long-range interactions gives evidence that even seemingly very complicated models may exhibit integrability, which is often indispensable to understand the dynamics of a system. There is a host of evidence that planar AdS/CFT correspondence may be one such system and several long-range spin chains have found applications in this string/gauge theory duality. This has already allowed to study many non-perturbative aspects of the duality, see \cite{chapTwist}. Moreover, methods based on integrability have allowed to conjecture the spectral equations of the planar AdS/CFT correspondence, see \cite{chapTrans}.
\section*{Acknowledgements}
I would like to thank Till Bargheer, Niklas Beisert, Marc Magro and Christoph Sieg for pointing out some shortcomings in the initial manuscript. The author is supported by a STFC postdoctoral fellowship.
\appendix
\newpage
\section{Three-Loop Hamiltonian of a generic long-range spin chain with \texorpdfstring{$\mathfrak{gl}(n)$}{gl(n)} symmetry algebra}
\begin{table}[h!]
\begin{align}
Q_2(\lambda) & =
              ( \PTerm{} - \PTerm{0} ) \nn\\[3mm]
           &      + \alpha_{0}(\lambda)\, ( - 3 \PTerm{} + 4 \PTerm{0} - \PTerm{0,1,0} ) \nn\\[3mm]
           & + \alpha_{0}(\lambda)^2 
               ( 20 \PTerm{} - 29 \PTerm{0} + 10 \PTerm{0,1,0} - \PTerm{0,1,2} - \PTerm{2,1,0} + \PTerm{0,2,1} + \PTerm{1,0,2} \nn \\ 
           &   \qquad - \PTerm{0,1,2,1,0} ) \nn \\
           & + \tfrac{i}{2} \alpha_{1}(\lambda)\,
               ( - 6 \PTerm{0,1} +6 \PTerm{1,0} + \PTerm{0,1,2,1} - \PTerm{1,2,1,0}
                   + \PTerm{0,1,0,2} - \PTerm{0,2,1,0} ) \nn \\
           & + \tfrac{1}{2} \beta_{2,3}(\lambda)\,
               ( -4 \PTerm{} + 8 \PTerm{0} - 2 \PTerm{0,1} - 2 \PTerm{1,0} - 2 \PTerm{0,2} \nn \\
           &   \qquad  - 2 \PTerm{0,1,2} - 2 \PTerm{2,1,0}+ 2 \PTerm{0,2,1} + 2 \PTerm{1,0,2}  \nn \\
           &   \qquad + \PTerm{0,1,2,1} + \PTerm{1,2,1,0} + \PTerm{0,1,0,2} + \PTerm{0,2,1,0}- 2 \PTerm{1,0,2,1} ) \nn \\
           & + i\epsilon_{2,1}(\lambda)\, ( \PTerm{1,0,2} - \PTerm{0,2,1} ) \nn \\
           & + i\epsilon_{2,2}(\lambda)\, (-\PTerm{0,1,2,1} + \PTerm{1,2,1,0}
                   + \PTerm{0,1,0,2} - \PTerm{0,2,1,0} )\nn\\[3mm]
           & +\Op\{\lambda^3\}\nn
\end{align}
\caption{Normalised Hamiltonian up to third order.}
\label{tab:Q2}
\end{table}

%%%%%%%%%%%%%%%%%%%%%%%%%%%%%%%%%%%%%%%%
\phantomsection
\addcontentsline{toc}{section}{\refname}
\bibliography{longrange,chapters}

%bibliography generated by nb.bst v1.01 (C) 2003-2010 Niklas Beisert
\begin{thebibliography}{10}
\ifx\href\asklfhas\newcommand{\href}[2]{#2}\fi
\ifx\arxivref\asklfhas\newcommand{\arxivref}[2]{\href{http://arxiv.org/abs/#1}%
{#2}}\fi
\ifx\doiref\asklfhas\newcommand{\doiref}[2]{\href{http://dx.doi.org/#1}{#2}}\fi
\raggedright
\small
\parskip 0pt

\bibitem{Maldacena:1997re}
J.~M.~Maldacena,
\textit{``{The large N limit of superconformal field theories and
  supergravity}''},
\textsf{Adv.~Theor.~Math.~Phys.~2,~231~(1998)},
\texttt{\arxivref{hep-th/9711200}{hep-th/9711200}}.
%%CITATION = HEP-TH/9711200;%%

\bibitem{Minahan:2002ve}
J.~A.~Minahan and K.~Zarembo,
\textit{``{The Bethe-ansatz for N = 4 super Yang-Mills}''},
\textsf{\doiref{10.1088/1126-6708/2003/03/013}{JHEP~0303,~013~(2003)}},
\texttt{\arxivref{hep-th/0212208}{hep-th/0212208}}.
%%CITATION = HEP-TH/0212208;%%

\bibitem{Beisert:2003jj}
N.~Beisert,
\textit{``{The complete one-loop dilatation operator of N = 4 super Yang-Mills
  theory}''},
\textsf{\doiref{10.1016/j.nuclphysb.2003.10.019}{Nucl.~Phys.~B676,~3~(2004)}},
\texttt{\arxivref{hep-th/0307015}{hep-th/0307015}}.
%%CITATION = HEP-TH/0307015;%%

\bibitem{Beisert:2003tq}
N.~Beisert, C.~Kristjansen and M.~Staudacher,
\textit{``{The dilatation operator of N = 4 super Yang-Mills theory}''},
\textsf{\doiref{10.1016/S0550-3213(03)00406-1}{Nucl.~Phys.~B664,~131~(2003)}},
\texttt{\arxivref{hep-th/0303060}{hep-th/0303060}}.
%%CITATION = HEP-TH/0303060;%%

\bibitem{Beisert:2003jb}
N.~Beisert,
\textit{``{Higher loops, integrability and the near BMN limit}''},
\textsf{\doiref{10.1088/1126-6708/2003/09/062}{JHEP~0309,~062~(2003)}},
\texttt{\arxivref{hep-th/0308074}{hep-th/0308074}}.
%%CITATION = HEP-TH/0308074;%%

\bibitem{Beisert:2003ys}
N.~Beisert,
\textit{``{The su(2$|$3) dynamic spin chain}''},
\textsf{\doiref{10.1016/j.nuclphysb.2003.12.032}{Nucl.~Phys.~B682,~487~(2004)}%
},
\texttt{\arxivref{hep-th/0310252}{hep-th/0310252}}.
%%CITATION = HEP-TH/0310252;%%

\bibitem{Zwiebel:2005er}
B.~I.~Zwiebel,
\textit{``{N = 4 SYM to two loops: Compact expressions for the non- compact
  symmetry algebra of the su(1,1$|$2) sector}''},
\textsf{\doiref{10.1088/1126-6708/2006/02/055}{JHEP~0602,~055~(2006)}},
\texttt{\arxivref{hep-th/0511109}{hep-th/0511109}}.
%%CITATION = HEP-TH/0511109;%%

\bibitem{Zwiebel:2008gr}
B.~I.~Zwiebel,
\textit{``{Iterative Structure of the N=4 SYM Spin Chain}''},
\textsf{\doiref{10.1088/1126-6708/2008/07/114}{JHEP~0807,~114~(2008)}},
\texttt{\arxivref{0806.1786}{arxiv:0806.1786}}.
%%CITATION = 0806.1786;%%

\bibitem{chapHigher}
C.~Sieg,
\textit{``Review of AdS/CFT Integrability, Chapter I.2: The spectrum from
  perturbative gauge theory''},
\texttt{\arxivref{1012.3984}{arxiv:1012.3984}}.
%%CITATION = 1012.3984;%%

\bibitem{Inozemtsev:1989yq}
V.~I.~Inozemtsev,
\textit{``{On the connection between the one-dimensional $s=1/2$ Heisenberg
  chain and Haldane-Shastry model,}''},
\textsf{J.~Stat.~Phys.~59,~1143~(1990)}.

\bibitem{Beisert:2005wv}
N.~Beisert and T.~Klose,
\textit{``{Long-range gl(n) integrable spin chains and plane-wave matrix
  theory}''},
\textsf{J.~Stat.~Mech.~0607,~P006~(2006)},
\texttt{\arxivref{hep-th/0510124}{hep-th/0510124}}.
%%CITATION = HEP-TH/0510124;%%

\bibitem{Bargheer:2008jt}
T.~Bargheer, N.~Beisert and F.~Loebbert,
\textit{``{Boosting Nearest-Neighbour to Long-Range Integrable Spin Chains}''},
\textsf{\doiref{10.1088/1742-5468/2008/11/L11001}{J.~Stat.~Mech.~0811,~L11001~%
(2008)}},
\texttt{\arxivref{0807.5081}{arxiv:0807.5081}}.
%%CITATION = 0807.5081;%%

\bibitem{Bargheer:2009xy}
T.~Bargheer, N.~Beisert and F.~Loebbert,
\textit{``{Long-Range Deformations for Integrable Spin Chains}''},
\textsf{\doiref{10.1088/1751-8113/42/28/285205}{J.~Phys.~A42,~285205~(2009)}},
\texttt{\arxivref{0902.0956}{arxiv:0902.0956}}.
%%CITATION = 0902.0956;%%

\bibitem{chapChain}
J.~A.~Minahan,
\textit{``Review of AdS/CFT Integrability, Chapter I.1: Spin Chains in
  $\mathcal{N}$ = 4 SYM''},
\texttt{\arxivref{1012.3983}{arxiv:1012.3983}}.
%%CITATION = 1012.3983;%%

\bibitem{Berenstein:2002jq}
D.~E.~Berenstein, J.~M.~Maldacena and H.~S.~Nastase,
\textit{``{Strings in flat space and pp waves from N = 4 super Yang Mills}''},
\textsf{\doiref{10.1088/1126-6708/2002/04/013}{JHEP~0204,~013~(2002)}},
\texttt{\arxivref{hep-th/0202021}{hep-th/0202021}}.
%%CITATION = HEP-TH/0202021;%%

\bibitem{Eden:2006rx}
B.~Eden and M.~Staudacher,
\textit{``{Integrability and transcendentality}''},
\textsf{J.~Stat.~Mech.~0611,~P014~(2006)},
\texttt{\arxivref{hep-th/0603157}{hep-th/0603157}}.
%%CITATION = HEP-TH/0603157;%%

\bibitem{chapObserv}
C.~Kristjansen,
\textit{``Review of AdS/CFT Integrability, Chapter IV.1: Aspects of
  Non-planarity''},
\texttt{\arxivref{1012.3997}{arxiv:1012.3997}}.
%%CITATION = 1012.3997;%%

\bibitem{Tetelman}
M.~Tetelman,
\textsf{Sov.~Phys.~JETP~55,~306~(1981)}.

\bibitem{WadatiSogo}
K.~Sogo and M.~Wadati,
\textit{``{Boost Operator and Its Application to Quantum Gelfand-Levitan
  Equation for Heisenberg-Ising Chain with Spin One-Half}''},
\textsf{\doiref{10.1143/PTP.69.431}{Prog.~Theor.~Phys.~69 No.2,~431~(1983)}}.

\bibitem{Grabowski:1994rb}
M.~Grabowski and P.~Mathieu,
\textit{``{Integrability test for spin chains}''},
\textsf{\doiref{10.1088/0305-4470/28/17/013}{J.Phys.A~A28,~4777~(1995)}},
\texttt{\arxivref{hep-th/9412039}{hep-th/9412039}}.

\bibitem{Beisert:2004hm}
N.~Beisert, V.~Dippel and M.~Staudacher,
\textit{``{A novel long range spin chain and planar N = 4 super Yang-
  Mills}''},
\textsf{\doiref{10.1088/1126-6708/2004/07/075}{JHEP~0407,~075~(2004)}},
\texttt{\arxivref{hep-th/0405001}{hep-th/0405001}}.
%%CITATION = HEP-TH/0405001;%%

\bibitem{Ryzhov:2004nz}
A.~V.~Ryzhov and A.~A.~Tseytlin,
\textit{``{Towards the exact dilatation operator of N = 4 super Yang- Mills
  theory}''},
\textsf{\doiref{10.1016/j.nuclphysb.2004.07.037}{Nucl.~Phys.~B698,~132~(2004)}%
},
\texttt{\arxivref{hep-th/0404215}{hep-th/0404215}}.
%%CITATION = HEP-TH/0404215;%%

\bibitem{Beisert:2007hz}
N.~Beisert, T.~McLoughlin and R.~Roiban,
\textit{``{The Four-Loop Dressing Phase of N=4 SYM}''},
\textsf{\doiref{10.1103/PhysRevD.76.046002}{Phys.~Rev.~D76,~046002~(2007)}},
\texttt{\arxivref{0705.0321}{arxiv:0705.0321}}.
%%CITATION = 0705.0321;%%

\bibitem{Beisert:2005tm}
N.~Beisert,
\textit{``{The su(2$|$2) dynamic S-matrix}''},
\textsf{Adv.~Theor.~Math.~Phys.~12,~945~(2008)},
\texttt{\arxivref{hep-th/0511082}{hep-th/0511082}}.
%%CITATION = HEP-TH/0511082;%%

\bibitem{chapSProp}
P.~Vieira and D.~Volin,
\textit{``Review of AdS/CFT Integrability, Chapter III.3: The dressing
  factor''},
\texttt{\arxivref{1012.3992}{arxiv:1012.3992}}.
%%CITATION = 1012.3992;%%

\bibitem{Agarwal:2005jj}
A.~Agarwal and G.~Ferretti,
\textit{``{Higher charges in dynamical spin chains for SYM theory}''},
\textsf{\doiref{10.1088/1126-6708/2005/10/051}{JHEP~0510,~051~(2005)}},
\texttt{\arxivref{hep-th/0508138}{hep-th/0508138}}.
%%CITATION = HEP-TH/0508138;%%

\bibitem{Beisert:2008qy}
N.~Beisert,
\textit{``{The su(2$|$3) Undynamic Spin Chain}''},
\textsf{\doiref{10.1143/PTPS.177.1}{Prog.~Theor.~Phys.~Suppl.~177,~1~(2009)}},
\texttt{\arxivref{0807.0099}{arxiv:0807.0099}}.
%%CITATION = 0807.0099;%%

\bibitem{Belitsky:2005bu}
A.~V.~Belitsky, G.~P.~Korchemsky and D.~Mueller,
\textit{``{Integrability of two-loop dilatation operator in gauge theories}''},
\textsf{\doiref{10.1016/j.nuclphysb.2005.11.015}{Nucl.~Phys.~B735,~17~(2006)}},
\texttt{\arxivref{hep-th/0509121}{hep-th/0509121}}.
%%CITATION = HEP-TH/0509121;%%

\bibitem{chapABA}
M.~Staudacher,
\textit{``Review of AdS/CFT Integrability, Chapter III.1: Bethe Ans\"atze and
  the R-Matrix Formalism''},
\texttt{\arxivref{1012.3990}{arxiv:1012.3990}}.
%%CITATION = 1012.3990;%%

\bibitem{chapSMat}
C.~Ahn and R.~I.~Nepomechie,
\textit{``Review of AdS/CFT Integrability, Chapter III.2: Exact world-sheet
  S-matrix''},
\texttt{\arxivref{1012.3991}{arxiv:1012.3991}}.
%%CITATION = 1012.3991;%%

\bibitem{Beisert:2007jv}
N.~Beisert and D.~Erkal,
\textit{``{Yangian Symmetry of Long-Range gl(N) Integrable Spin Chains}''},
\textsf{\doiref{10.1088/1742-5468/2008/03/P03001}{J.~Stat.~Mech.~0803,~P03001~%
(2008)}},
\texttt{\arxivref{0711.4813}{arxiv:0711.4813}}.
%%CITATION = 0711.4813;%%

\bibitem{chapYang}
A.~Torrielli,
\textit{``Review of AdS/CFT Integrability, Chapter VI.2: Yangian Algebra''},
\texttt{\arxivref{1012.4005}{arxiv:1012.4005}}.
%%CITATION = 1012.4005;%%

\bibitem{Beisert:2008cf}
N.~Beisert and F.~Loebbert,
\textit{``{Open Perturbatively Long-Range Integrable gl(N) Spin Chains}''},
\textsf{Adv.~Sci.~Lett.~2,~261~(2009)},
\texttt{\arxivref{0805.3260}{arxiv:0805.3260}}.
%%CITATION = 0805.3260;%%

\bibitem{Serban:2004jf}
D.~Serban and M.~Staudacher,
\textit{``{Planar N = 4 gauge theory and the Inozemtsev long range spin
  chain}''},
\textsf{\doiref{10.1088/1126-6708/2004/06/001}{JHEP~0406,~001~(2004)}},
\texttt{\arxivref{hep-th/0401057}{hep-th/0401057}}.
%%CITATION = HEP-TH/0401057;%%

\bibitem{Haldane:1987gg}
F.~D.~M.~Haldane,
\textit{``{Exact Jastrow-Gutzwiller resonating valence bond ground state of the
  spin 1/2 antiferromagnetic Heisenberg chain with 1/r**2 exchange}''},
\textsf{\doiref{10.1103/PhysRevLett.60.635}{Phys.~Rev.~Lett.~60,~635~(1988)}}.
%%CITATION = PRLTA,60,635;%%

\bibitem{Shastry:1987gh}
B.~Sriram~Shastry,
\textit{``{Exact solution of an S = 1/2 Heisenberg antiferromagnetic chain with
  long ranged interactions}''},
\textsf{\doiref{10.1103/PhysRevLett.60.639}{Phys.~Rev.~Lett.~60,~639~(1988)}}.
%%CITATION = PRLTA,60,639;%%

\bibitem{Lieb:1968zza}
E.~H.~Lieb and F.~Y.~Wu,
\textit{``{Absence of Mott transition in an exact solution of the short-range,
  one-band model in one dimension}''},
\textsf{\doiref{10.1103/PhysRevLett.20.1445}{Phys.~Rev.~Lett.~20,~1445~(1968)}%
}.
%%CITATION = PRLTA,20,1445;%%

\bibitem{Rej:2005qt}
A.~Rej, D.~Serban and M.~Staudacher,
\textit{``{Planar N = 4 gauge theory and the Hubbard model}''},
\textsf{\doiref{10.1088/1126-6708/2006/03/018}{JHEP~0603,~018~(2006)}},
\texttt{\arxivref{hep-th/0512077}{hep-th/0512077}}.
%%CITATION = HEP-TH/0512077;%%

\bibitem{chapTwist}
L.~Freyhult,
\textit{``Review of AdS/CFT Integrability, Chapter III.4: Twist states and the
  cusp anomalous dimension''},
\texttt{\arxivref{1012.3993}{arxiv:1012.3993}}.
%%CITATION = 1012.3993;%%

\bibitem{chapTrans}
V.~Kazakov and N.~Gromov,
\textit{``Review of AdS/CFT Integrability, Chapter III.7: Hirota Dynamics for
  Quantum Integrability''},
\texttt{\arxivref{1012.3996}{arxiv:1012.3996}}.
%%CITATION = 1012.3996;%%

\end{thebibliography}
\bibliographystyle{nb}

\end{document}